\input{aipcheck}

\documentclass[
  ,final            
  ]
  {aipproc}

\layoutstyle{8x11double}

\usepackage{amsmath}
\usepackage[pdftex,dvipsnames]{color}

\begin{document}

\title{\vspace{-2.5cm}{\small Published in: M. Nakagawa and S. Luding (eds.), Powders and Grains 2009: Proceedings of the 6th International\\[-0.3cm] Conference on Micromechanics of Granular Media, American Institute of Physics (Melville, New York, 2009), p. 859-862}\\[1.5cm] Fractal Substructures due to Fragmentation and Reagglomeration}


\author{D. E. Wolf}{
  address={Department of Physics and CeNIDE, University of
        Duisburg-Essen, 47048 Duisburg, Germany}
}
\author{T. P\"oschel}{
  address={CBI and Center of Excellence EAM, Universit\"at 
Erlangen-N\"urnberg, 91058 Erlangen, Germany} 
}
\author{T. Schwager}{
  address={Charit\'e, Augustenburger Platz 1, 13353 Berlin, Germany} 
}
\author{A. Weuster}{
  address={Department of Physics and CeNIDE, University of
        Duisburg-Essen, 47048 Duisburg, Germany}
}
\author{L. Brendel}{
  address={Department of Physics and CeNIDE, University of
        Duisburg-Essen, 47048 Duisburg, Germany}
}

\begin{abstract}
  Cohesive powders form agglomerates that can be very porous. Hence
  they are also very fragile. Consider a process of complete
  fragmentation on a characteristic length scale $\ell$, where the
  fragments are subsequently allowed to settle under gravity. If this
  fragmentation-reagglomeration cycle is repeated sufficiently often,
  the powder develops a fractal substructure with robust statistical
  properties. The structural evolution is discussed for two different
  models: The first one is an off-lattice model, in which a fragment does
  not stick to the surface of other fragments that have already
  settled, but rolls down until it finds a locally stable position.
  The second one is a simpler lattice model, in which a fragment
  sticks at first contact with the agglomerate of fragments that have
  already settled. Results for the fragment size distribution are
  shown as well. One can distinguish scale invariant dust and
  fragments of a characteristic size. Their role in the process of
  structure formation will be addressed.

\end{abstract}

\maketitle


\section{Introduction}

Most particles attract each other, be it by
van-der-Waals forces or by microscopic liquid menisci at the contacts,
but often this attraction is so weak compared to other forces acting
on the particles that it may 
safely be neglected, when explaining the physical behaviour of
granular matter. There are notable exceptions, however, where
attractive forces are decisive. In this paper we consider one such
case, in which the attractive force between two particles is much larger
than their weight. This is true for nanopowders
\cite{Maedler}, but also for larger particles under
microgravity \cite{Blum}, or for wet sand \cite{Herminghaus}. 

Such a granular medium with adhesion between the particles can sustain
very high porosity in spite of gravity. This can be easily shown: Take
two equal 
glass containers, one empty and the other one at most half filled with
dry sand.  Porosity is known to be about
0.36 (random dense packing). Then about 10 volume percent of
water is kneaded into the sand. One gets a smooth dough. By and by
this dough is fragmented by means of a fork into crumbs which are
poured into the empty container. One finds the filling height
increased. Then the crumb assembly is again fragmented with the fork
and filled back into the original container. Again the filling height
increases. Repeating this procedure a few times leads to a sand
packing with porosity of about 70 percent. Thermophoretic aerosol deposits can
even have porosities as high as 99 percent.

In this paper we report on the structural properties of a porous
assembly of adhesive particles obtained by repeated fragmentation and
reagglomeration.  We show that a fractal substructure forms and that
the fragment size distribution is very broad. To obtain statistically
significant results for the structure one has to
consider systems with more than one million particles for many
fragmentation-reagglomeration cycles.  This is beyond the capability
of Molecular Dynamics simulations. Therefore, two simplified 
two-dimensional models were studied. 

\begin{figure}[t]
  \centering
\includegraphics[width=0.93\columnwidth,clip]{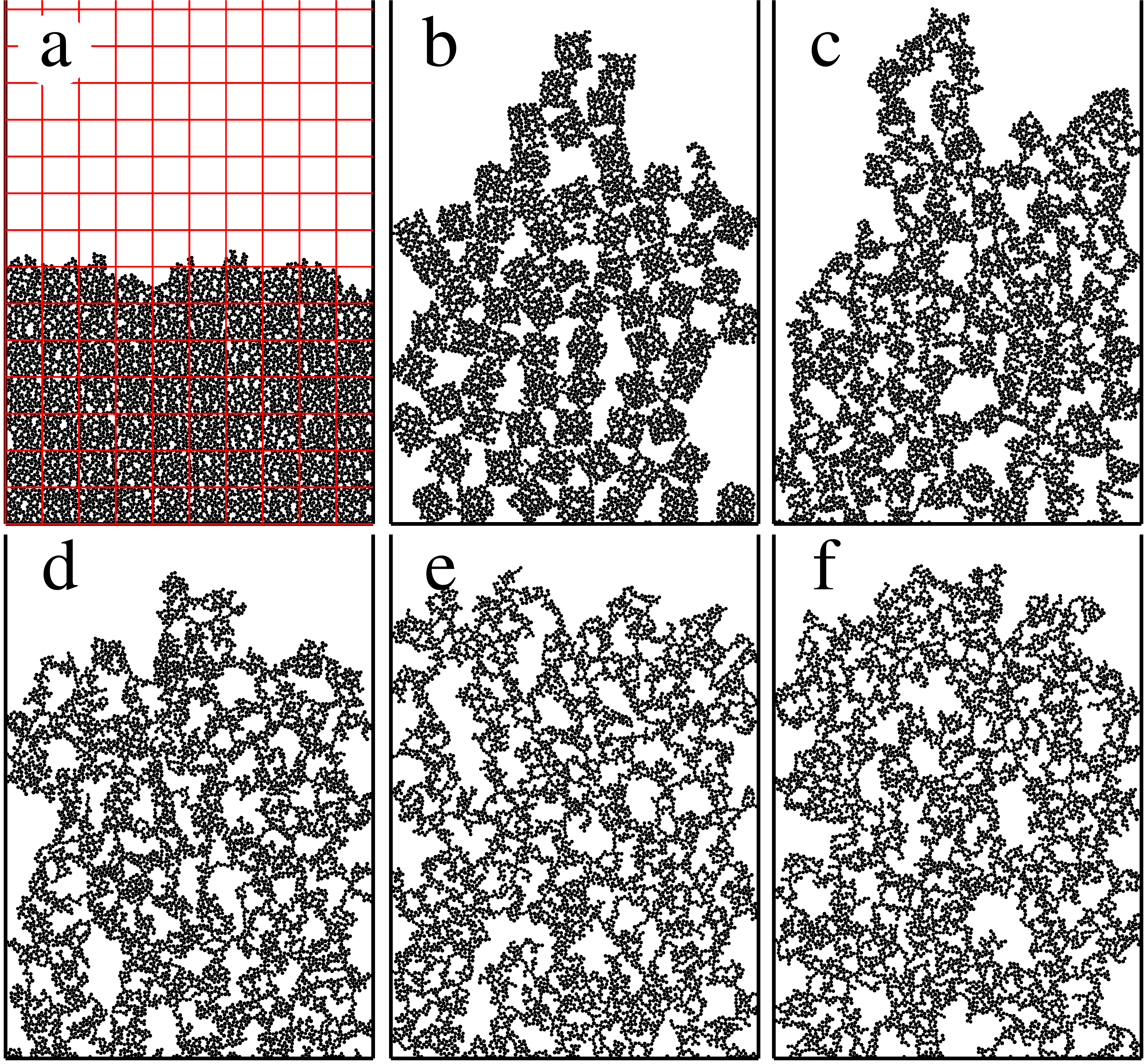}
  \caption{Evolution of the packing in the off-lattice model.
    a) Initial packing generated by random sequential sedimentation
    \cite{VisscherBolsterli:1972}. The packing is cut by a square mesh into
    fragments ($\ell=20$); b) The fragments are considered as rigid
    bodies and deposited  
    (1$^{st}$ generation). Again the packing is cut by the square mesh 
    (here not shown); c) the fragments are deposited again (2$^{nd}$
    generation), and so on; 
    d) 3$^{rd}$ generation; e) 4$^{th}$ generation; f) 120$^{th}$ generation.}
  \label{fig:generationsA}
\end{figure}

\begin{figure}[t]
  \centering
  \includegraphics[width=0.8\columnwidth]{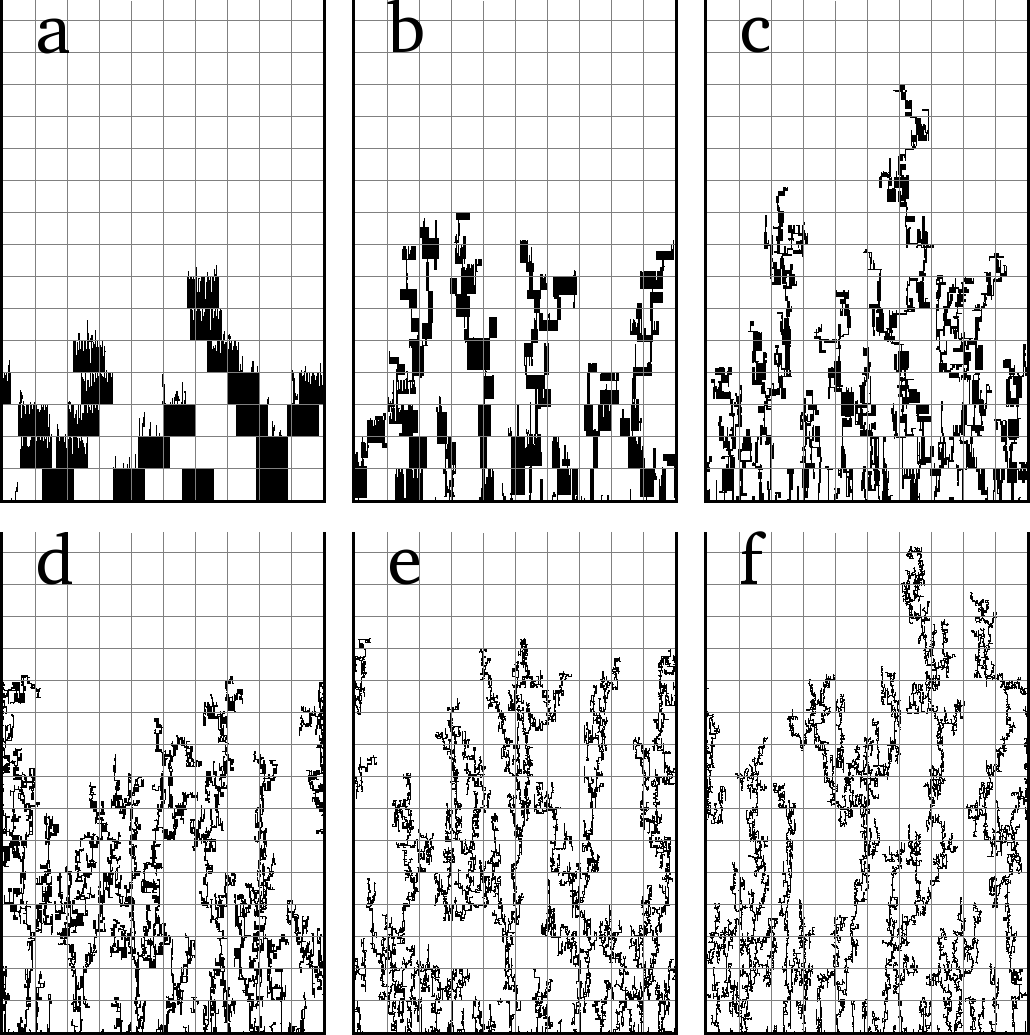}
  \caption{Evolution of the packing in the lattice model.
    a)1$^{st}$ generation. The initial packing was cut by the
    indicated square mesh into 
    fragments ($\ell=32$); b) 4$^{th}$ generation; c) 10$^{th}$ generation; 
    d) 20$^{th}$ generation; e) 40$^{th}$ generation; f) 100$^{th}$
    generation.} 
  \label{fig:generationsB}
\end{figure}

The first one, an off-lattice model \cite{Schwager08}, is a
generalization of a model for the sequential deposition of
non-adhesive spherical particles under the influence of gravity
\cite{VisscherBolsterli:1972,jullien93b,jullien87b,jullien88,baumann93,baumann95}.
The second one is a lattice model, generalizing the model of ballistic
deposition \cite{Vold59,Meakin86}. In both models the
following procedure is repeated many times: First the agglomerate is
cut with a square mesh into portions. The linear mesh size $\ell$ can
be viewed as the typical scale of the fragmentation process. A portion
may consist of several disconnected fragments. The models differ in the way 
these fragments then settle under gravity. They do so 
as rigid bodies without taking adhesion
forces with other particles into account. This is justified, if the
flakes are sufficiently large, so that their weight exceeds the
adhesive force between the particles. After this reassembly of the
fragments the agglomerate is cut again with the square mesh, and so
on, see Fig.~\ref{fig:generationsA} and Fig.~\ref{fig:generationsB}.
In the following, lengths are given in units of the average particle
radius (off-lattice model) respectively the lattice constant (lattice
model), masses in units of the particle mass, and time as number of 
fragmentation-reagglomeration cycles.

\section{The off-lattice model \cite{Schwager08}}

In this model the reagglomeration of the fragments is simulated in the
following way: Each fragment starts at a random position with a random
orientation well above the already deposited material (the
configuration of which is regarded as frozen in). Following gravity,
it moves downwards until it touches the bottom of the container or
contacts another already deposited particle. Then it rolls down as a
rigid body, again following gravity, until the vertical projection of
its center of mass falls in between two points of contact (with the
container or previously deposited particles).   

We simulated up to 3 million particles represented by discs with a
narrow size distribution (10\% variance).  The initial state was a
densely packed agglomerate. In each fragmentation-reagglomeration
cycle the filling height increases. Asymptotically, the powder adopts
a very porous, statistically invariant structure, which is robust with
respect to fragmentation at scale $\ell$ and subsequent gravitational
settling of the fragments.
  
\begin{figure}
  \centering
\includegraphics[width=0.8\columnwidth]{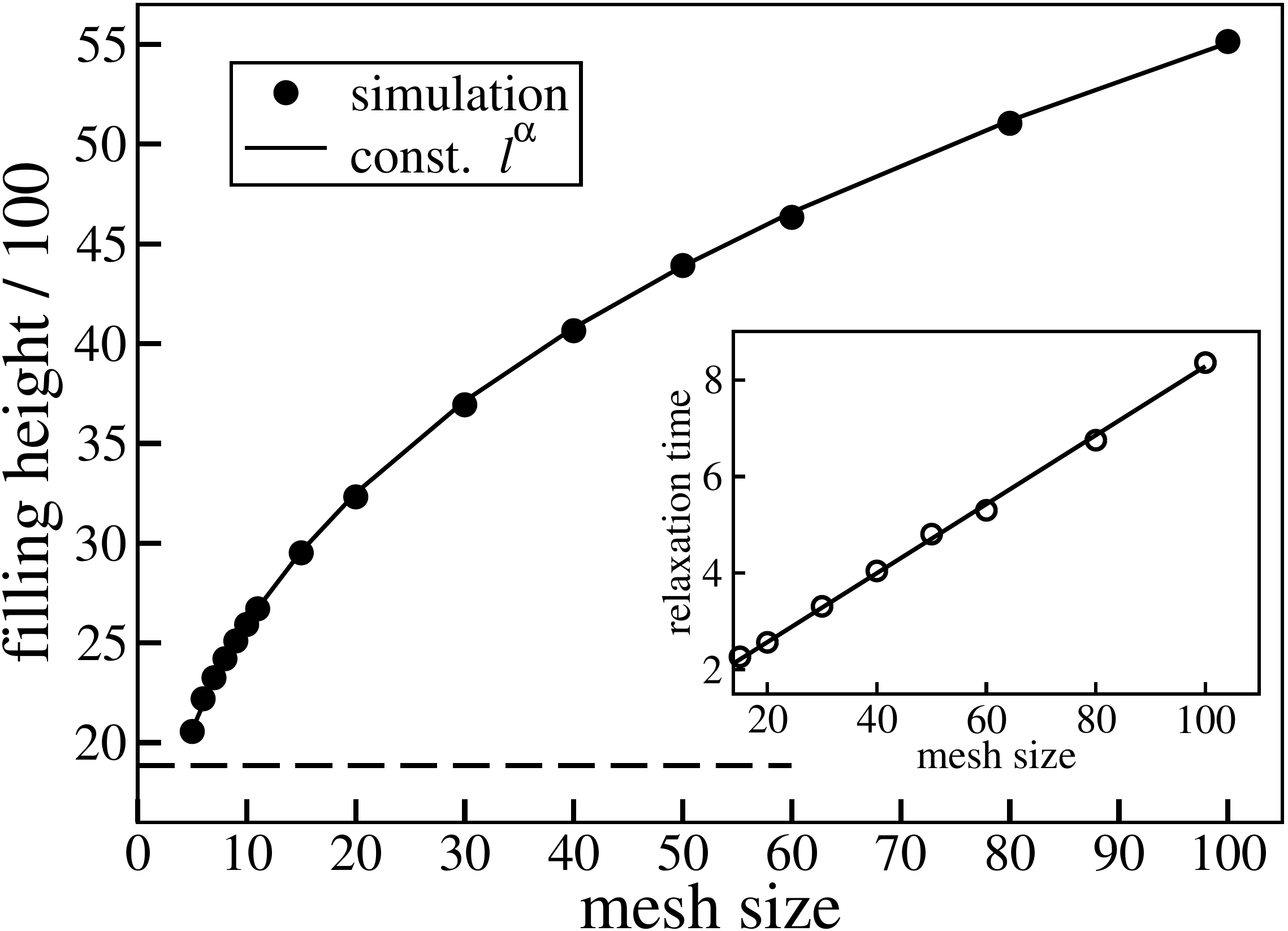}  
  \caption{The asymptotic filling heigth $h_{\infty}$ grows as a
  power law $h_\infty(\ell)\sim\ell^\alpha$ with mesh size $\ell$. The
  full line shows the best fit, $\alpha=0.327$. Inset: The relaxation
  time $n_0(\ell)$ increases linearly with mesh size.} 
  \label{fig:height_bottom}
\end{figure}

The filling height $h_{n}$ at iteration step
$n$ approaches the asymptotic height,
$h_{\infty}$, exponentially with a relaxation time, $n_0$. 
The inset of Fig.~\ref{fig:height_bottom} shows that
\begin{equation}
n_0(\ell) \propto \ell^z \quad \text{with} \quad z=1.
\end{equation}
For the asymptotic filling height, a power law 
\begin{equation}
h_{\infty}(\ell) \propto \ell^{\alpha} \quad \text{with} \quad \alpha = 0.327
\end{equation}
gives a very good fit (see Fig.~\ref{fig:height_bottom}). 
This implies that the number of portions cut from the steady state
configuration of a system of width $W$ scales like $N_\text{p} =
h_{\infty}W/\ell^2 \propto 
\ell^{\alpha-2}$. Consequently, the mass of a $\ell\times\ell$-portion
has a fractal dimension $d_\text{f}$,
\begin{equation} 
\frac{M}{N_\text{p}} \propto \ell^{d_\text{f}} \quad \text{with} \quad
d_\text{f} = 2-\alpha = 1.67 \pm 0.03. 
\label{eq:d_f}
\end{equation}

\begin{figure}[t]
   \centering
\includegraphics[width=0.8\columnwidth,bb=15 34 719 531]{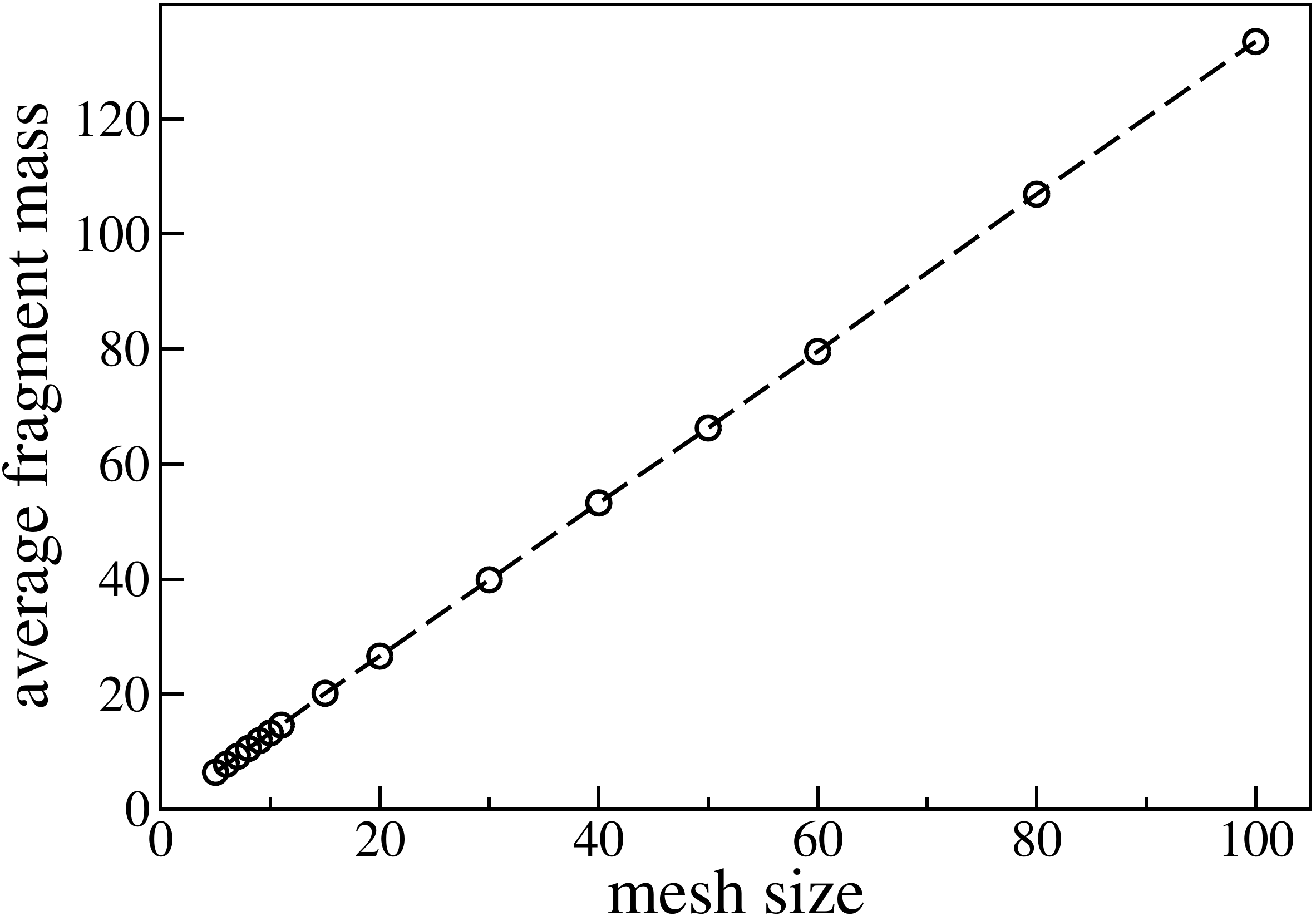}
  \caption{Average fragment mass
    as a function of the mesh size $\ell$ in the steady
    state of the off-lattice model.} 
  \label{fig:mean_fragment_size_over_mesh}
\end{figure}

In the steady state the number of fragments per portion is determined
by the fact that the fragmentation process cuts on average as many
particle contacts as will be reestablished by the agglomeration
process. Since every fragment, when settling, creates two new contacts,
the number of fragments per portion must be proportional to the number
of contacts cut at the boundary of a mesh cell,
\begin{equation}
\frac{N_\text{f}}{N_\text{p}} \propto \ell^{d_\text{f}-1}.
\label{eq:fragments_per_portion}  
\end{equation}
This is confirmed by Fig. \ref{fig:mean_fragment_size_over_mesh},
which shows that $M/N_\text{f}\propto \ell$.

\begin{figure}[t]
  \centering
\includegraphics[width=0.8\columnwidth,angle=0,clip]{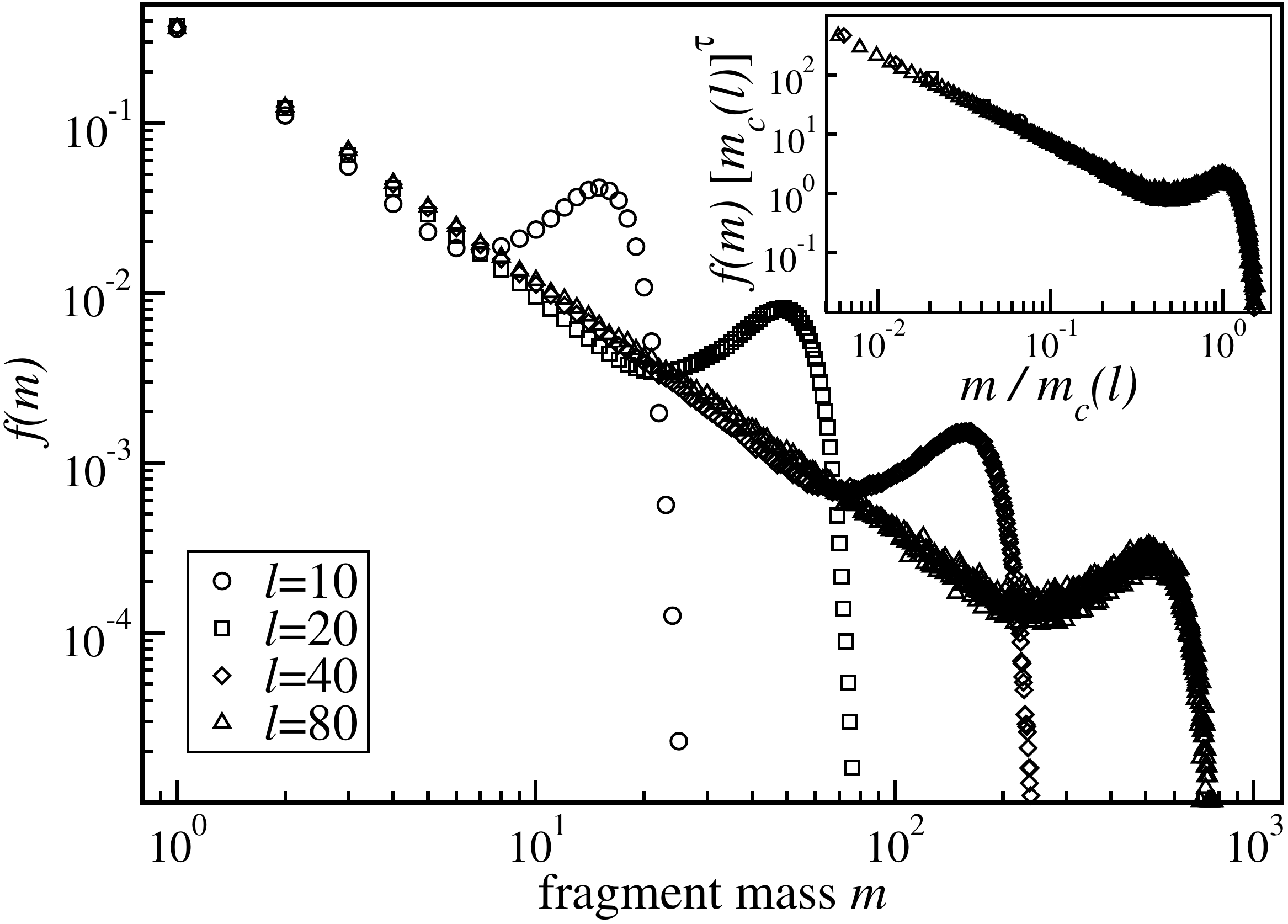} 
  \caption{Normalized fragment mass distribution for different
  mesh sizes $\ell$. $f(m)$ is the number of fragments of mass $m$
  divided by the total number of fragments for a given $\ell$. Inset:
  Data collapse using $m_\text{c}\propto \ell^{d_\text{f}}$ with
  $d_\text{f}=1.695$, and $\tau=1.41$.}   
  \label{fig:cluster_sizes}
\end{figure}

A detailed understanding of the fragment properties is provided by
the distribution of fragment masses, shown in Fig.
\ref{fig:cluster_sizes}. Two types of fragments must be distinguished,
large chunks at the upper end of the mass spectrum with a
characteristic size $m_\text{c}$, and scale invariant dust responsible
for the power law part that is cut off by $m_\text{c}$. Comparing the
mass distributions for different mesh sizes $\ell$ shows, that they
can approximately be written in the form
\begin{equation}
  \label{eq:cluster_size}
  f(m,\ell) = m^{-\tau} \tilde{f}\left(\frac{m}{m_\text{c}(\ell)}\right)\ ,
\end{equation}
where the scaling function $\tilde{f}(x)$ is constant for $x\ll 1$, goes through a
maximum at $x=1$, and has an approximately Gaussian tail for $x\gg 1$. 
\begin{figure}[t]
  \centering
\includegraphics[width=0.8\columnwidth,clip]{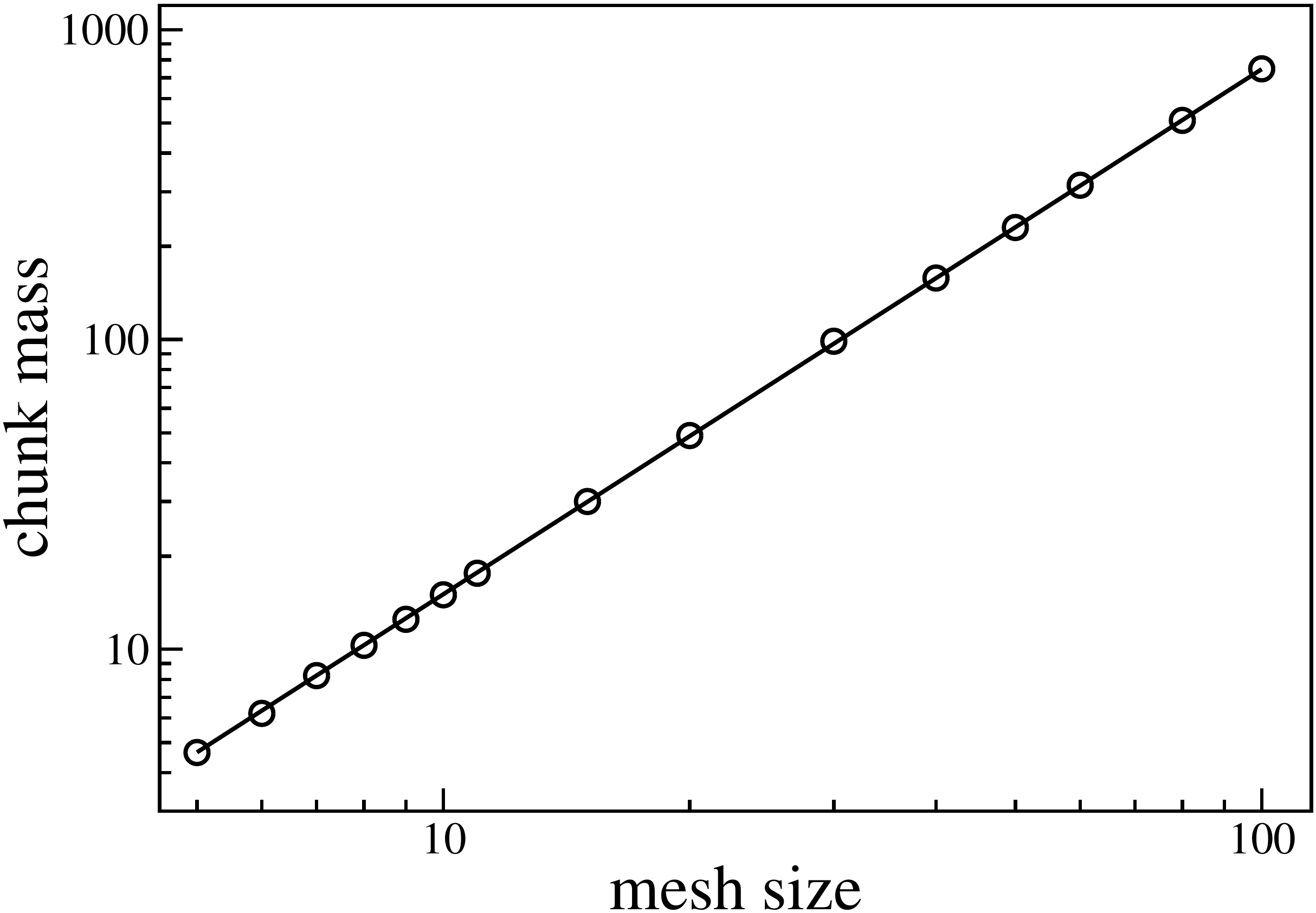}
  \caption{Chunk mass $m_c$ as a function of mesh size. Slope of straight line
  is 1.695.}
  \label{fig:chunk_mass}
\end{figure}
The typical mass $m_\text{c}$ of the chunks has a power-law dependence on the
mesh size, $m_\text{c} = 0.304~\ell^{1.695}$ (Fig.~\ref{fig:chunk_mass}), the
exponent being in good agreement with the value of $d_\text{f}$,
Eq.~\eqref{eq:d_f}.

The fractal chunks are only a tiny fraction of the total number of fragments,
$N_{\rm f}$. This follows from the fact that the width of the
chunk-distribution is proportional to $m_\text{c}$. Therefore, the
number of chunks, $N_{\rm c}$, divided by the number of fragments,
$N_{\rm f}$, decreases with increasing mesh size like
\begin{equation}
\frac{N_\text{c}}{N_\text{f}} \propto m_\text{c}^{1-\tau} \propto
\ell^{d_\text{f}(1-\tau)}. 
\label{eq:f_chunks}
\end{equation}

The fractal dimension of the chunks and the dust exponent $\tau$ are
not independent of each other but obey the scaling relation
\begin{equation}
d_\text{f}(2-\tau)=1\ .
\label{eq:scaling_relation}
\end{equation}
The reason is the following: As the chunk mass $m_\text{c} \propto
\ell^{d_\text{f}}$ scales in the same way as the total
mass per portion, $M/N_\text{p}$, the number of chunks per portion
cannot depend on $\ell$. Using (\ref{eq:fragments_per_portion}) and
(\ref{eq:f_chunks}) one concludes that 
\begin{equation}
\text{const.} = \frac{N_\text{c}}{N_\text{p}} \propto
\ell^{d_\text{f}(1-\tau)}\ell^{d_\text{f}-1},
\end{equation}
which proves (\ref{eq:scaling_relation}).
For $d_\text{f}=1.695$ this implies $\tau=1.41$. These values lead to
an excellent data 
collapse for the fragment mass distributions (see
Fig.~\ref{fig:cluster_sizes} (inset)).  

We have seen, that the overwhelming number of fragments are dust
particles, apart from a vanishing fraction (\ref{eq:f_chunks}) of
chunks.  This explains, why the mass (essentially mass
of chunks) per fragment (essentially per dust particle) has nothing to
do with the fractal dimension, but is proportional to $\ell$ (see
Fig.~\ref{fig:mean_fragment_size_over_mesh}) .

\section{The lattice model}

In order to explore how universal the fractal substructure is, we
studied another fragmentation-reagglomeration model. Here all
particles live on a square lattice (of width $W$ and height $H$) with
lateral periodic boundary conditions, which in each cycle is
subdivided into square cells of size $\ell\times \ell$. Then,
traversing from bottom to top and each row from left to right, the
content of each cell is considered. Disconnected fragments therein are
identified and - without changing their orientation - subsequently
deposited into an (initially empty) $W\times H$-container at a
randomly chosen horizontal position. Deposition means here, that the
fragment sinks vertically until it forms the first \emph{vertical}
contact with the container bottom or a previously deposited cluster.
At this point it stops moving.  In contrast to the ballistic
deposition model \cite{Meakin86} horizontal contacts have no effect
for the deposition process, but they become as sticky as any other
contact, once the fragment is at rest. As the fragments do not roll
down until they form a second contact, the structure, see
Fig.\ref{fig:generationsB}, is more treelike than in
Fig.\ref{fig:generationsA}.

In the following, we show results for a system with $N=2^{26}$
particles (occupied lattice sites) in a container of width $W=4096$.
The mesh size ranges from $\ell=16,\dots,256$. Data are averaged over
6 independent runs, starting from initial conditions of randomly
deposited single particles. As for the off-lattice model the system
approaches a steady state in relaxation time $n_0 \propto \ell$, i.e.\ 
the exponent $z$ is 1. 
\begin{figure}[t]
  \centering
  \includegraphics[width=0.8\columnwidth]{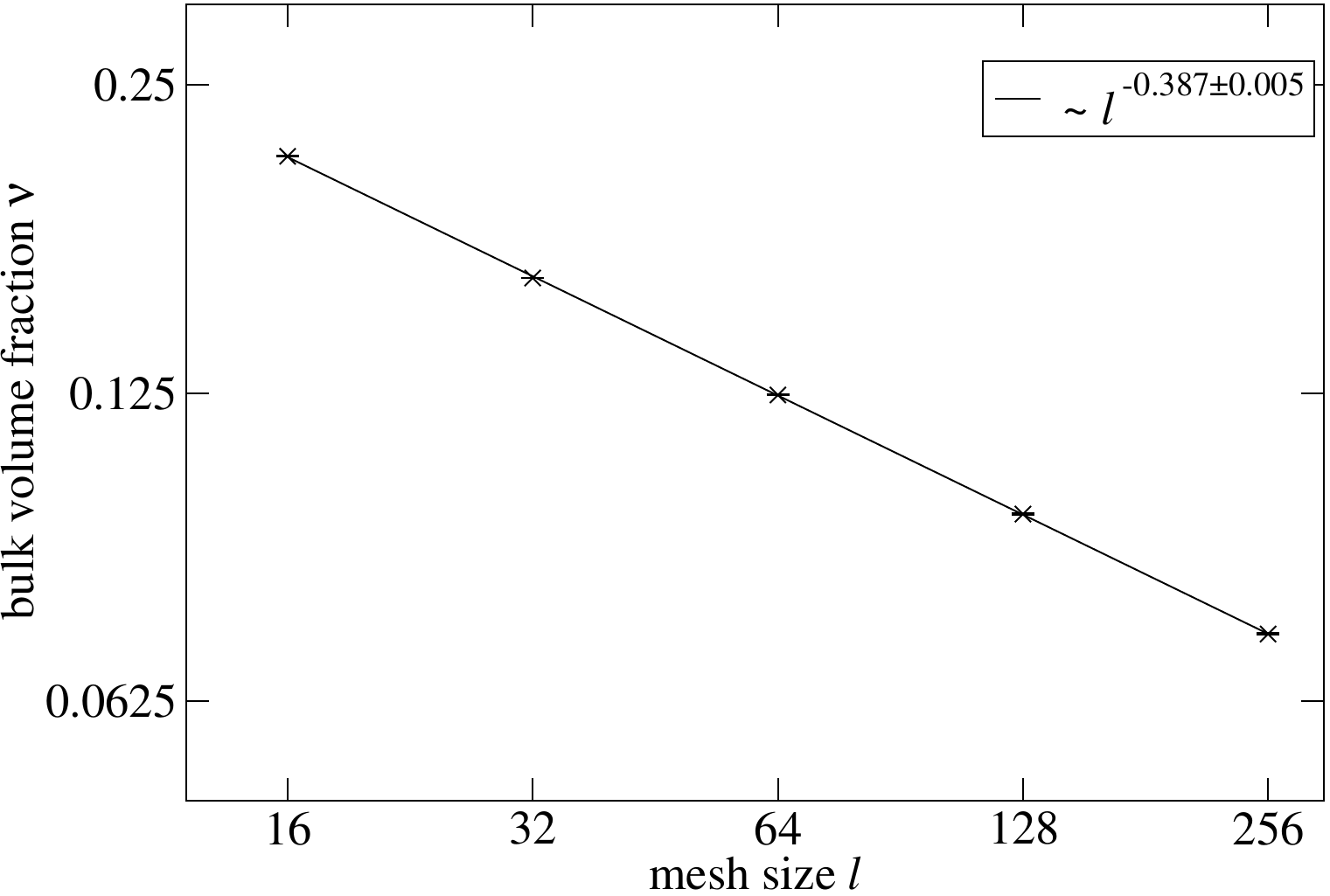}
  \caption{The steady state bulk density $\nu$ as a function of the
    mesh size $\ell$ for the lattice model.}
  \label{fig:RCD_nu_of_a}
\end{figure}
\begin{figure}
  \centering
  \includegraphics[width=0.8\columnwidth]{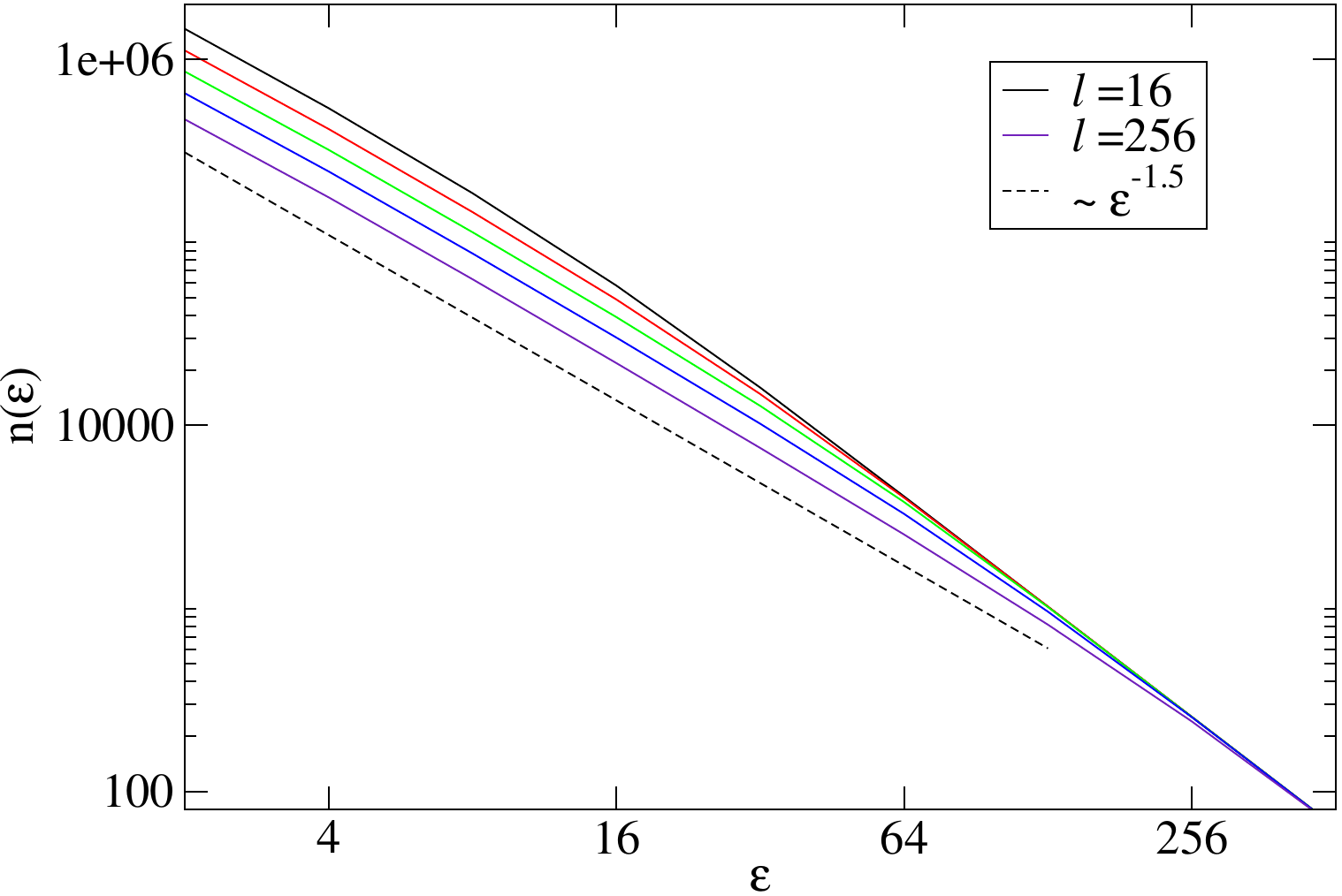}
  \caption{The box counting method ($\epsilon$ being the box size)
    applied to the steady state bulk structure of the lattice model:
    Below the mesh size $\ell$, a fractal dimension of about 
    $1.5$ prevails.}
  \label{fig:RCD_n_of_eps}
\end{figure}

Figure \ref{fig:RCD_nu_of_a} shows that the bulk volume fraction $\nu$
in the steady state is a power law of the mesh size $\ell$ with a
non-trivial exponent: $\nu\propto
\ell^{d_\text{f}-2}=\ell^{-0.387\pm0.005}$.
This indicates again a fractal substructure up to scale $\ell$ which
is confirmed in Fig.\ref{fig:RCD_n_of_eps}. 
The fractal dimension $d_\text{f}=1.5\pm 0.1$ obtained from the box
counting method should be more reliable than the one extracted from the global
solid fraction, Fig.\ref{fig:RCD_nu_of_a}, because of the crossover
from the fractal structure on small scales to the homogeneous density
on scales larger than $\ell$. For the lattice model the fractal
dimension seems to be a bit smaller than for the off-lattice
model. Lattice anisotropy has a similar effect on the fractal
behaviour of DLA \cite{Meakin}.

\subsection{Acknowledgement}
We would like to thank I. Goldhirsch for fruitful discussions. This
work was partially supported by the German Research Foundation 
within the collaborative research center SFB 445 ``Nanoparticles from
the gas phase'' and by the German-Israeli Foundation by grant
no. I-795-166.10/2003.


\begin{thebibliography}{99}
\bibitem{Maedler} L.~M\"adler, A.~A.~Lall, and S.~K.~Friedlander,
  Nanotechnology {\bf 17}, 4783 (2006).
\bibitem{Blum} J.~Blum, G.~Wurm, Annu.Rev.Astron.Astr. {\bf 46}, 21 (2008).
\bibitem{Herminghaus} S.~Herminghaus, Adv.Phys. {\bf 54}, 221 (2005).
\bibitem{Schwager08} T.~Schwager, D.~E.~Wolf, and T.~P\"oschel, 
  Phys. Rev. Lett. {\bf 100}, 218002 (2008).
\bibitem{VisscherBolsterli:1972} W.~M.~Visscher and M.~Bolsterli,
  Nature {\bf 239}, 504 (1972).
\bibitem{jullien93b} R.~Jullien, P.~Meakin, and A.~Pavlovitch,
  Europhys. Lett. {\bf 22}, 523 (1993).
\bibitem{jullien87b} R.~Jullien and P.~Meakin, Europhys. Lett. {\bf 4}, 1385
  (1987).
\bibitem{jullien88} R.~Jullien and P.~Meakin, Europhys. Lett. {\bf 6}, 629
  (1988). 
\bibitem{baumann93} G.~Baumann, E.~Jobs, and D.~E.~Wolf, Fractals {\bf 1}, 767
  (1993). 
\bibitem{baumann95} G.~Baumann, I.~M.~J\'anosi, and D.~E.~Wolf, Phys. Rev. E
  {\bf 51}, 1879 (1995).
\bibitem{Vold59} M.~J.~Vold, J. Colloid Sci. {\bf 14}, 168 (1959).
\bibitem{Meakin86} P.~Meakin, P.~Ramanlal, L.~M.~Sander, R.~C.~Ball,
  Phys. Rev. A {\bf 34}, 5091 (1986).
\bibitem{Meakin} P.~Meakin, in {\it Phase Transitions and Critical
    Phenomena}, edited by C.~Domb and J.~L.~Lebowitz (Academic Press,
    London, 1988), vol. 12, p.336.
\end{thebibliography}
\end{document}